\documentclass[aps,pra,amsmath,amsfonts,amssymb,twocolumn,superscriptaddress,showpacs]{revtex4-1}
\usepackage{amssymb,amsmath,amsfonts}
\usepackage{graphicx,psfrag,color}
\usepackage{dcolumn}
\usepackage{bbm}
\usepackage{mathbbol}
\usepackage{braket}
\usepackage{verbatim}
\usepackage{hyperref}
\hypersetup{
  colorlinks=true,
  linkcolor=blue,
  citecolor=blue,
  urlcolor=blue
}

\newcommand*{\balancecolsandclearpage}{%
  \close@column@grid
  \clearpage
  \twocolumngrid
}

\begin{document}
\flushbottom

\title{Continuous-time quantum walks on a defective lattice: Boosting the spreading \\ of delocalized states through Parrondo’s strategy} 

\author{Jo\~ao V. Cordeiro}
\affiliation{Departamento de F\'isica, Universidade do Estado de Santa Catarina, 89219-710, Joinville, Santa Catarina, Brazil}

\author{Eduardo I. Duzzioni}
\affiliation{Departamento de F\'isica, Universidade Federal de Santa Catarina, 88040-900, Florian\'opolis, Santa Catarina, Brazil} 

\author{Edgard P. M. Amorim}
\email{edgard.amorim@udesc.br} 
\affiliation{Departamento de F\'isica, Universidade do Estado de Santa Catarina, 89219-710, Joinville, Santa Catarina, Brazil}

\date{\today}

\begin{abstract}
We investigate the quantum transport of delocalized states in continuous-time quantum walks (CTQWs) on a one-dimensional lattice containing a single defect. The defect is modeled by assigning complex-valued hopping amplitudes to the edges that connect the site corresponding to the mean position of the delocalized initial state to its nearest neighbors. We find that this single defective site is sufficient to enhance the ballistic spreading of an initially Gaussian wave packet. Extending these results, we implement a time-dependent alternation protocol between two distinct defect configurations, each individually yielding poor propagation of the state. The combination of these two unfavorable configurations improves the transport efficiency of the quantum walker, revealing a manifestation of Parrondo’s paradox in CTQWs with delocalized initial states. This study provides insights into the role of complex-phase defects and time-dependent protocols in CTQWs, demonstrating that the interplay between quantum interference and graph engineering can effectively enhance quantum transport in discrete lattices.
\end{abstract}


\maketitle

\section{Introduction} \label{sec:intro}

Quantum walks (QWs), the quantum analogs of classical random walks, have emerged as fundamental models in quantum information processing and quantum computing, providing a versatile framework for the development of quantum algorithms. The concept was initially proposed in 1993 by Aharonov \textit{et al.} in a discrete-time formulation~\cite{aharonov1993quantum} and later extended to a continuous-time framework in 1998 by Farhi and Gutmann~\cite{farhi1998quantum}. In contrast to classical random walks, whose variance grows linearly with time, $\sigma^2\propto t$, quantum walks exhibit ballistic spreading, characterized by a quadratic growth of the variance, $\sigma^2\propto t^2$~\cite{kempe2003quantum,venegas2012quantum}. This enhanced spreading originates from coherent quantum interference and allows the walker to explore the position space more efficiently. The resulting propagation advantage motivated early applications of quantum walks to quantum search algorithms, where they provide a quadratic speedup in hitting times relative to classical counterparts, enabling the identification of an item among $N$ possibilities in $O(\sqrt{N})$ steps~\cite{shenvi2003quantum,tulsi2008faster,portugal2013quantum}. More broadly, they have since been employed to model a wide range of physical processes, including photosynthetic energy transport~\cite{engel2007evidence,mohseni2008environment} and universal quantum computation~\cite{childs2009universal,lovett2010universal}. Furthermore, they have been realized across diverse experimental platforms~\cite{wang2013physical,flamini2019photonic}.

The transport properties of quantum walks are sensitive to both dynamical and spatial variations~\cite{vieira2013dynamically,vieira2014entangling,konno2010localization,wojcik2012trapping,li2013position,zhang2014one,endo2014one,teles2021localization,li2015single,ximenes2024parrondo}. When quantum coins are chosen randomly over time, the walker exhibits diffusive behavior~\cite{vieira2013dynamically}, whereas spatial disorder along the lattice leads to Anderson localization~\cite{vieira2014entangling}. Even a single lattice defect may be sufficient to trap or localize the quantum state~\cite{konno2010localization,wojcik2012trapping,li2013position,zhang2014one,endo2014one,teles2021localization}, although in some cases it may also enhance its propagation~\cite{li2015single,ximenes2024parrondo}. Since the propagation efficiency of quantum walks depends on improving their ballistic spreading, strategies that promote faster walker propagation can be regarded as successful, in contrast to those that localize or trap the state~\cite{keating2007localization}.

In this context, an intriguing link arises between game theory and quantum mechanics through Parrondo’s effect in quantum phenomena~\cite{lai2020review}. Parrondo’s effect describes apparently paradoxical situations in which losing strategies or adverse effects combine to produce a winning outcome~\cite{harmer2002review,shu2014beyond}. This effect has been explored across several fields. For instance, it has been used to explain complex adaptations in biological systems~\cite{jean2021,cheong2019}, to optimize crop rotations in agriculture~\cite{chaitanya2023}, to analyze complex networks~\cite{cheong2016}, and to design routing strategies~\cite{ankit2024}. Turning to quantum walks, one may wonder whether alternating over time between two individual lattice defects, each leading to poor spreading, would also result in a weak outcome. This question was recently addressed by Ximenes \textit{et al.}~\cite{ximenes2024parrondo}. In their study, the authors modified the hopping amplitudes at a single site and identified two distinct hopping configurations, each of which individually leads to poor spreading. They then demonstrated Parrondo’s effect, showing that alternating between these two configurations during the walk enhances the walker’s propagation, thereby outperforming the case without hopping modulation.

Parrondo's effect has been investigated extensively in discrete-time quantum walks~\cite{chandrashekar2011Parrondo,flitney2012quantum,li2013quantum,rajendran2018implementing,lai2020parrondo,pires2020parrondo,jan2020experimental,panda2022generating,trautmann2022parrondo,mittal2024parrondo} and was subsequently demonstrated in continuous-time quantum walks (CTQWs) by Ximenes \textit{et al.}~\cite{ximenes2024parrondo}. Their analysis, however, was restricted to a single localized initial state and real hopping amplitudes. By considering superpositions of position states instead, one can further explore phase effects arising from complex hopping amplitudes. Several studies have revealed novel features of quantum walks initialized in delocalized states~\cite{tregenna2003controlling,valcarcel2010tailoring,romanelli2010distribution,zhang2016creating,orthey2017asymptotic,orthey2019connecting,orthey2019weak}, including solitonlike behavior~\cite{ghizoni2019trojan} and potential applications to high-fidelity state transfer~\cite{vieira2021quantum,engster2024high}. These states were also realized experimentally on a variety of platforms~\cite{cardano2015quantum,su2019experimental,derrico2020two}, where engineered phase shifters can implement complex hopping amplitudes. Chiral quantum walks introduced in Ref.~\cite{lu2016chiral} and subsequent works~\cite{khalique2021controlled,chaves2023why,bottarelli2023quantum} demonstrated directional control of quantum transport driven by complex hopping phases. In the present work, we combine delocalized initial states and complex hopping amplitudes within a Parrondo-type alternation protocol to probe their impact on wave-packet spreading.

Here, we investigate the dynamics of a one-dimensional CTQW on a defective lattice starting from a Gaussian state, employing a Parrondo-based strategy to enhance the spreading of the walker. The defect is introduced through complex hopping amplitudes between the site corresponding to the mean position of the Gaussian initial state and its nearest neighbors. We systematically explore the influence of both the real and imaginary components of the hopping amplitude, thereby identifying regions associated with good or poor spreading regimes relative to the defect-free case. Subsequently, we implement Parrondo’s strategy, showing a significant enhancement in the spreading of the delocalized state.

This article is organized as follows. Section~\ref{sec:2} presents the mathematical formalism of the CTQW model on a defective lattice, considering a Gaussian distribution as the initial state. Section~\ref{sec:3} discusses the spreading behavior of the delocalized state under complex defects and the implementation of Parrondo’s strategy. Finally, Sec.~\ref{sec:4} summarizes the main conclusions.

\section{Mathematical formalism}\label{sec:2}

In the following sections, we present the theoretical framework of CTQWs, including the incorporation of a defect in the one-dimensional lattice, the delocalized initial state considered here, and the alternation protocol used to implement Parrondo’s strategy.

\subsection{Continuous-time quantum walks}\label{sec:2.1}

The CTQW model explored here describes the motion of a quantum particle on an infinite line composed of discrete sites. The quantum particle can hop only between neighboring sites, i.e., from site $j$ to $j\pm 1$. A quantum state $\ket{\psi}$ associated with the particle evolves via the Schr\"{o}dinger equation,
\begin{equation}
i\hbar\frac{\partial}{\partial t}\ket{\psi(t)}=\hat{H}\ket{\psi(t)},
\label{Scheq}
\end{equation}
with the Hamiltonian for a CTQW on a homogeneous lattice given by
\begin{equation}
\hat{H}=\epsilon\sum_j \ket{j}\bra{j}-\gamma\sum_j (\ket{j-1}\bra{j}+\ket{j+1}\bra{j}),
\label{H0}
\end{equation}
where $\epsilon$ denotes the constant on-site potential energy and $\gamma$ is the hopping amplitude between neighboring sites.

After we obtain $\ket{\psi(t)}$, the probability of finding the particle at site $j$ is
\begin{equation}
P_j(t) = \left|\braket{j|\psi(t)}\right|^2,
\label{Pjt}
\end{equation}
and to quantify the spreading of the particle, we consider the standard deviation,
\begin{equation}
\sigma(t)=\sqrt{\braket{j^2}-\braket{j}^2},
\label{sigmat}
\end{equation}
where $\braket{j^n}=\bra{\psi(t)}j^n\ket{\psi(t)}$ denotes the $n$th moment of the position distribution. The quantum walker exhibits ballistic spreading, meaning that the standard deviation grows linearly in time, $\sigma\propto t$, in contrast to the classical diffusive scaling $\sigma\propto\sqrt{t}$. For $t\gg 1$, the quantum walker reaches the long-time regime~\cite{orthey2019connecting}, where the slope or the time derivative of the standard deviation,
\begin{equation}
\alpha=\frac{d\sigma(t)}{dt},
\label{slope}
\end{equation} 
quantifies the walker's long-time spreading rate.

\subsection{CTQW on the defective lattice}\label{sec:2.2}

\begin{figure}[!ht]
\includegraphics[width=\linewidth]{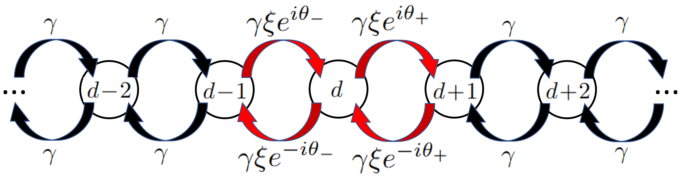}
\caption{Schematic representation of a CTQW on a defective lattice. The hopping amplitudes from site $d$ to its nearest neighbors $d\pm1$ are $\gamma \xi e^{\pm i\theta_\pm}$, whereas all other hopping amplitudes are real and equal to $\gamma$. Two scenarios are considered: (1) both phases are equal, $\theta_-=\theta_+=\theta$, and (2) the phases have opposite signs, $\theta_-=\theta$ and $\theta_+=-\theta$. Throughout this study, we set $\epsilon=0$ from $\hat{H}$ in Eq.~\eqref{H_d} and take $d=0$ for convenience.}
\label{Fig1}
\end{figure}

Previous studies investigated the transport of a quantum particle in one-dimensional CTQWs on lattices containing a single defect~\cite{li2015single,ximenes2024parrondo}. Such a defect can be modeled by modifying the hopping amplitudes between a specific site $j=d$ and its neighboring sites $d\pm 1$. We therefore consider the following Hamiltonian: 
\begin{equation}
\hat{H}_d=\hat{H}+\gamma\big(\hat{V}_d+\hat{V}_d^\dagger\big),
\label{H_d}
\end{equation}
where $\hat{V}_d^\dagger$ is the Hermitian conjugate of $\hat{V}_d$ and the defect operator is defined as
\begin{equation}
\hat{V}_d=(1-\xi e^{-i\theta_-})\ket{d-1}\bra{d}+(1-\xi e^{i\theta_+})\ket{d+1}\bra{d}.
\label{V_d}
\end{equation}
This formulation implies that the hoppings from site $d$ to its nearest neighbors $d\pm1$ are scaled by dimensionless complex factors $\xi e^{\pm i\theta_\pm}$, while the reverse hoppings from sites $d\pm1$ to $d$ are multiplied by their corresponding complex conjugates. For $\theta_-\neq\theta_+$, the model introduces distinct phases in the left and right hopping amplitudes. Finally, the Hamiltonian $\hat{H}_d$ reduces to $\hat{H}$ when $\xi=1$ and $\theta_-=\theta_+=0$, whereas $\xi=0$ completely decouples site $d$, trapping the probability amplitude of the quantum walker at that position. Figure~\ref{Fig1} illustrates the defective lattice model.

\subsection{Delocalized initial state}\label{sec:2.3}

In this work, we consider an initial quantum state represented as a superposition of site basis states,
\begin{equation}
\ket{\psi_0}=\sum_j \psi(j)\ket{j},
\label{Psi0}
\end{equation}
where $\psi(j)$ is a distribution function constrained by the normalization condition $\sum_j \left|\psi(j)\right|^2 = 1$. Instead of the delta function $\psi(j)=\delta_{j,0}$, which represented a localized state in previous studies~\cite{li2015single,ximenes2024parrondo}, we consider a Gaussian distribution,
\begin{equation}
\psi(j)=Ae^{-\frac{j^2}{4\sigma_0^2}},
\label{psij}
\end{equation}
where $|\psi(j)|^2$ represents a Gaussian profile centered at the origin and $\sigma_0$ denotes the initial standard deviation of the distribution. For sufficiently delocalized states ($\sigma_0 \geq 1$), the normalization constant $A$ is well approximated by $(2\pi\sigma_0^2)^{-1/4}$. 

It is worth emphasizing that complex-weighted hopping amplitudes play a significant role in the transport properties of delocalized states by inducing chiral propagation, as shown in earlier works~\cite{lu2016chiral,khalique2021controlled,chaves2023why,bottarelli2023quantum} and further discussed in Sec.~\ref{sec:3}.

\subsection{Parrondo protocol}

The protocol implements the quantum analog of Parrondo’s paradox by alternating the system dynamics between two distinct defective configurations, each corresponding to a poor-spreading regime. In this protocol, we consider two Hamiltonians, $\hat{H}^{A}_d$ and $\hat{H}^{B}_d$, obtained from Eq.~\eqref{H_d} using two different parameter sets:
\begin{equation}
(\xi^{A},\theta_{+}^{A},\theta_{-}^{A}),\quad (\xi^{B},\theta_{+}^{B},\theta_{-}^{B}).
\end{equation}
If we take here $\gamma=1$, each half cycle of the protocol evolves the system under one of the Hamiltonians, $\hat{H}^{A}_d$ or $\hat{H}^{B}_d$, for a fixed time interval $\pi/\omega$, thereby defining the time-dependent Hamiltonian: 
\begin{equation}
\hat{H}_{AB}(t) =
\begin{cases}
\hat{H}^{A}_d, & 0 \le t < \pi/\omega,\\
\hat{H}^{B}_d, & \pi/\omega \le t < 2\pi/\omega,
\end{cases}
\label{H_p}
\end{equation}
where $\omega$ is the angular frequency corresponding to a full cycle of period $T=2\pi/\omega$, and then $\hat{H}_{AB}(t+T)=\hat{H}_{AB}(t)$. The dynamics proceeds as follows. The initial quantum state first evolves under $\hat{H}^{A}_d$, after which the resulting state evolves under $\hat{H}^{B}_d$. This sequence is then repeated, and the successive alternations determine the long-time spreading behavior of the walker.

\section{Results and Discussion}\label{sec:3}

We first analyze the spreading of a Gaussian state in continuous space, which serves as a reference for comparison with the discrete-lattice simulations. Next, we investigate how a CTQW on a defective lattice, initialized in a Gaussian state, behaves relative to the localized case. Finally, we map the parameter space to identify three representative alternation configurations that enhance propagation under Parrondo’s strategy.

\subsection{Spreading of a Gaussian state}

Although the evolution in CTQWs involves solving the Schrödinger equation using a discrete Laplacian operator~\cite{wong2016Laplacian}, we examine the continuous case for comparison with the numerical results. Consider a one-dimensional free-particle wave packet given by the distribution function $\phi(x)$, i.e., a continuous-position version of Eq.~\eqref{psij}. For convenience, we set $\gamma=1/(2m)$ and $\hbar=1$ in the Schrödinger equation. The Hamiltonian becomes
\begin{equation}
\hat{H}=-\gamma\frac{\partial^2}{\partial x^2};
\label{Hfree}
\end{equation}
then, the evolved state is given by 
\begin{equation}
\ket{\phi(t)}=e^{-i\hat{H}t}\ket{\phi(0)}.
\label{phit}
\end{equation}
The evolution of a free Gaussian wave packet can be found in many quantum mechanics textbooks \cite{cohen2020,nouredine2009,mcintyre2012}. A Gaussian wave packet $\phi(x)$ exhibits ballistic spreading, with an asymptotic spreading rate that decreases as the initial width increases. In the limit $\sigma_0\rightarrow0$, however, the continuum probability density $|\phi(x)|^2$ becomes singular and fails to reproduce the localized-state limit. 
This issue can be addressed by introducing an effective width $\sigma_{\rm eff}=\sigma_0/\mathrm{Erf}\!\left(\sqrt{\pi/2}\,\sigma_0\right)$, where $\mathrm{Erf}(z)=\frac{2}{\sqrt{\pi}}\int_0^z e^{-u^2}\,du$ denotes the error function~\cite{orthey2017asymptotic}. Accordingly, we replace the bare Gaussian by the renormalized initial state,
\begin{equation}
    \phi_c(x,0)=\left(2\pi\sigma_{\rm eff}^2\right)^{-1/4}
    \exp\!\left[-\frac{x^2}{4\sigma_{\rm eff}^2}\right],
\end{equation}
whose Fourier transform is also Gaussian in momentum space. Since the evolution operator is diagonal in the momentum representation, the time-evolved wave function is given by
\begin{equation}
\phi_c(x,t)=\frac{1}{\sqrt{2\pi}}\int_{-\infty}^{\infty}\phi_c(p,0)\,e^{-i\gamma t p^2}e^{ipx}\,dp,
\end{equation}
which can be evaluated analytically, yielding
\begin{equation}
\phi_c(x,t)\!=\!\frac{\sigma_{\rm eff}^{1/2}}{[2\pi(\sigma_{\rm eff}^2\!+\!i\gamma t)^2]^{1/4}}
\exp\!\left[-\frac{x^2}{4(\sigma_{\rm eff}^2\!+\!i\gamma t)}\right].
\end{equation}
It follows that the probability density $|\phi_c(x,t)|^2$ remains Gaussian, centered at the origin, with a time-dependent standard deviation
\begin{equation}
\sigma_c(t)=\sqrt{\sigma_{\rm eff}^2+\gamma^2\frac{t^2}{\sigma_{\rm eff}^2}}.
\label{csigmat}
\end{equation}
For $t\gg 1$, $\sigma_c(t)\approx \gamma t/\sigma_{\rm eff}$, yielding the corrected asymptotic spreading rate of a Gaussian state
\begin{equation}
\alpha_c\approx\frac{\gamma}{\sigma_0}\mathrm{Erf}\left(\sqrt{\frac{\pi}{2}}\sigma_0\right),
\label{alphacorr}
\end{equation}
instead of the uncorrected asymptotic spreading rate
\begin{equation}
\alpha_a\approx\gamma/\sigma_0.
\label{alphaasym}
\end{equation}
This correction leaves the broad-packet limit unchanged since $\mathrm{Erf}(\sqrt{\pi/2}\,\sigma_0)\to1$ for $\sigma_0\gg1$, while for $\sigma_0\to0$ it gives $\sigma_{\rm eff}\to1/\sqrt{2}$ and hence $\alpha_c\to\sqrt{2}\gamma$, recovering the localized-state result. Figure~2 compares the spreading rates obtained from CTQW calculations with the analytical predictions given by $\alpha_c$ and $\alpha_a$.

\begin{figure}[!ht]
\includegraphics[width=\linewidth]{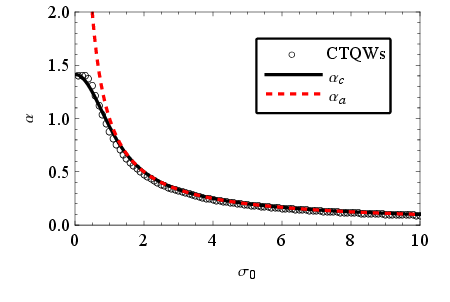}
\caption{Spreading rate $\alpha$ versus $\sigma_0$ for CTQWs evolved up to $\gamma t=1000$ (black open circles) from Gaussian initial states. The analytical expressions $\alpha_c$ [Eq.~\eqref{alphacorr}] (black line) and $\alpha_a$ [Eq.~\eqref{alphaasym}] (red dashed line) are shown for comparison, both evaluated at $\gamma=1$.}
\label{Fig2}
\end{figure}

\subsection{Gaussian states on the defective lattice}

In the previous section, we analyzed the long-time spreading dynamics of Gaussian wave packets governed by the Schrödinger equation as a function of their standard deviation $\sigma_0$. Although the evolution remains ballistic, the spreading rate $\alpha$ decreases asymptotically as $\sigma_0$ increases. In contrast to a $\delta$-localized state, whose momentum-space representation is uniform, a Gaussian state exhibits a Gaussian momentum distribution. As $\sigma_0$ increases, this distribution narrows, suppressing high-momentum components and thereby reducing the effective spreading rate. This behavior directly reflects the Heisenberg uncertainty principle: a larger initial spatial width implies a smaller momentum spread, limiting the contribution of fast-propagating components to the wave-packet evolution.

\begin{figure}[!ht]
\includegraphics[width=\linewidth]{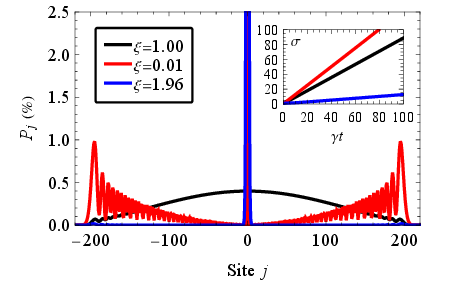}
\caption{Probability distribution $P_j$ at $\gamma t=100$ and standard deviation $\sigma$ as a function of $\gamma t$ (inset) for a CTQW with a Gaussian initial state of width $\sigma_0=1$. For $\xi=0.01$ (red line), the walk displays partial localization together with enhanced spreading relative to the defect-free case, $\xi=1$ (black line). By contrast, $\xi=1.96$ (blue line) produces strong localization and suppressed spreading. In all cases, $\theta_-=\theta_+=0$.}
\label{Fig3}
\end{figure}

From an experimental perspective, perfectly localized states are unattainable. Any realistic wave packet necessarily exhibits finite spatial and momentum uncertainties arising from intrinsic quantum limits and extrinsic classical imperfections introduced during state preparation. Although measurement-induced quantum disturbances are negligible compared to these classical contributions~\cite{peres2002quantum}, their combined effect results in delocalized initial conditions, which in turn reduce the spreading rate of the wave function. Consequently, delocalized states provide a more faithful representation of experimentally realizable quantum walkers, and understanding their transport behavior is essential for bridging theoretical models to practical implementations~\cite{wang2013physical,flamini2019photonic, cardano2015quantum,su2019experimental,derrico2020two}. In this setting, identifying conditions under which disorder, fluctuations, or lattice defects enhance spreading becomes particularly relevant since the performance of quantum walks ultimately depends on the walker’s propagation efficiency. At the same time, such mechanisms may induce partial localization without preventing spreading, making it necessary to assess whether the state becomes trapped near the defect, as shown in Fig.~\ref{Fig3}.

\begin{figure*}[!ht]
\includegraphics[width=\linewidth]{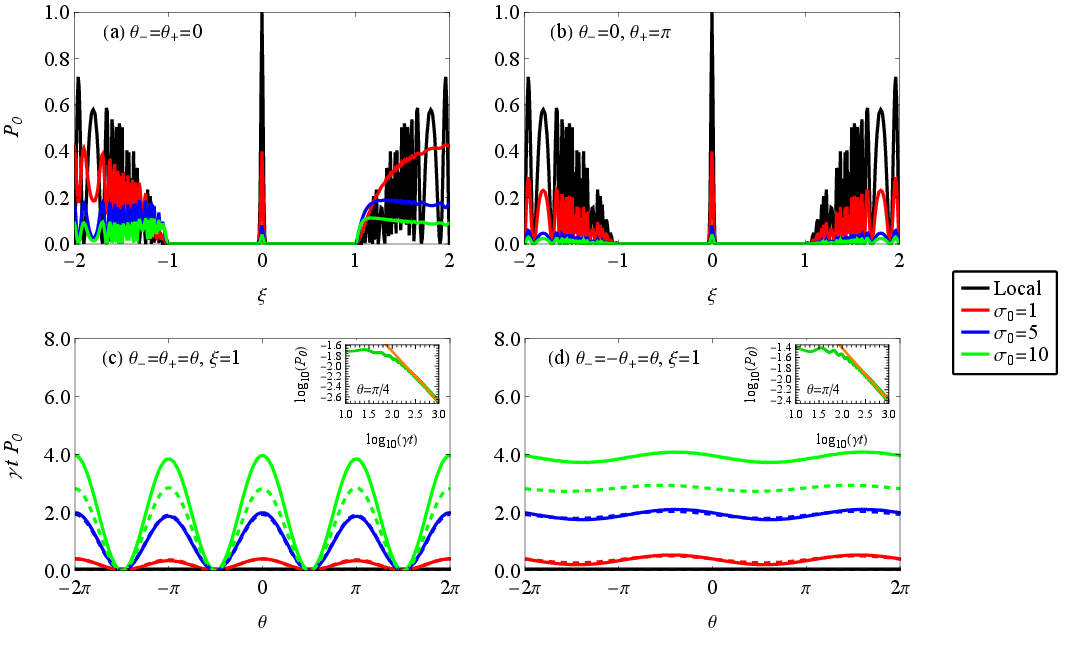}
\caption{
Probability $P_0$ at the defect site as a function of the hopping amplitude $\xi$ for (a) $\theta_-=\theta_+=0$ and (b) $\theta_-=0$ and $\theta_+=\pi$. (a) shows asymmetric $\xi$ dependence for Gaussian initial states, whereas (b) restores reflection symmetry through the relative phase difference. Rescaled probability $\gamma t\,P_0$ as a function of $\theta$ for (c) $\theta_-=\theta_+=\theta$ and (d) $\theta_-=-\theta_+=\theta$, both with $\xi=1$. Solid (dashed) lines correspond to evolution up to $\gamma t=1000$ ($\gamma t=100$). The insets in (c) and (d) show $\log_{10}(P_0)$ as a function of $\log_{10}(\gamma t)$ for $\sigma_0=10$ and $\theta=\pi/4$. The orange curves correspond to power-law fits for $\gamma t>500$, explicitly showing the asymptotic long-time behavior $P_0\propto(\gamma t)^{-1}$. The fully localized initial state (black lines) corresponds to $\ket{\psi_0}=\ket{0}$, while Gaussian initial states have $\sigma_0=1$ (red lines), $5$ (blue lines), and $10$ (green lines).}
\label{Fig4}
\end{figure*}

To establish a quantitative comparison between the spreading of CTQWs on a defective lattice, governed by $\hat{H}_d$ in Eq.~\eqref{H_d}, and that on a homogeneous lattice, driven by $\hat{H}$ in Eq.~\eqref{H0}, we use as a figure of merit the ratio $\sigma_d/\sigma$ between the corresponding standard deviations in the long-time regime ($\gamma t\gg 1$), following Ximenes \textit{et al.}~\cite{ximenes2024parrondo}. In addition, we evaluate the probability $P_0$ at the defect site to identify parameter regimes that lead to localization. 

Figure~\ref{Fig4} characterizes the influence of the defect parameters $(\xi,\theta_\pm)$ on localization in CTQWs initialized from Gaussian states with different widths. Figures \ref{Fig4}(a) and \ref{Fig4}(b) show the probability $P_0$ at the defect as a function of the hopping-amplitude ratio $\xi$ for $\theta_-=\theta_+=0$ [Fig. \ref{Fig4}(a)] and $\theta_-=0$ and $\theta_+=\pi$ [Fig. \ref{Fig4}(b)]. The nonmonotonic dependence of $P_0$ on $\xi$ reflects two distinct localization mechanisms. For small $|\xi|$, the defect site becomes weakly coupled to its neighbors, effectively isolating it and trapping probability, thereby locally suppressing transport. As $\xi \rightarrow 1$, the system recovers translational invariance, and $P_0$ converges to the defect-free value, indicating fully delocalized dynamics. For $|\xi|>1$, however, the defect bonds exceed the background coupling, creating a strong-bond defect that supports localized or quasilocalized modes. In this regime, a fraction of the amplitude oscillates between the defect and its nearest neighbors without propagating away, yielding a secondary increase in $P_0$. This revival of localization at large $|\xi|$, arising from strong-coupling-induced reflection, is closely analogous to bound-state formation in tight-binding lattices with a local on-site impurity~\cite{ashcroft1976solid,kittel2005introduction}.

\begin{figure*}[!ht]
\includegraphics[width=\linewidth]{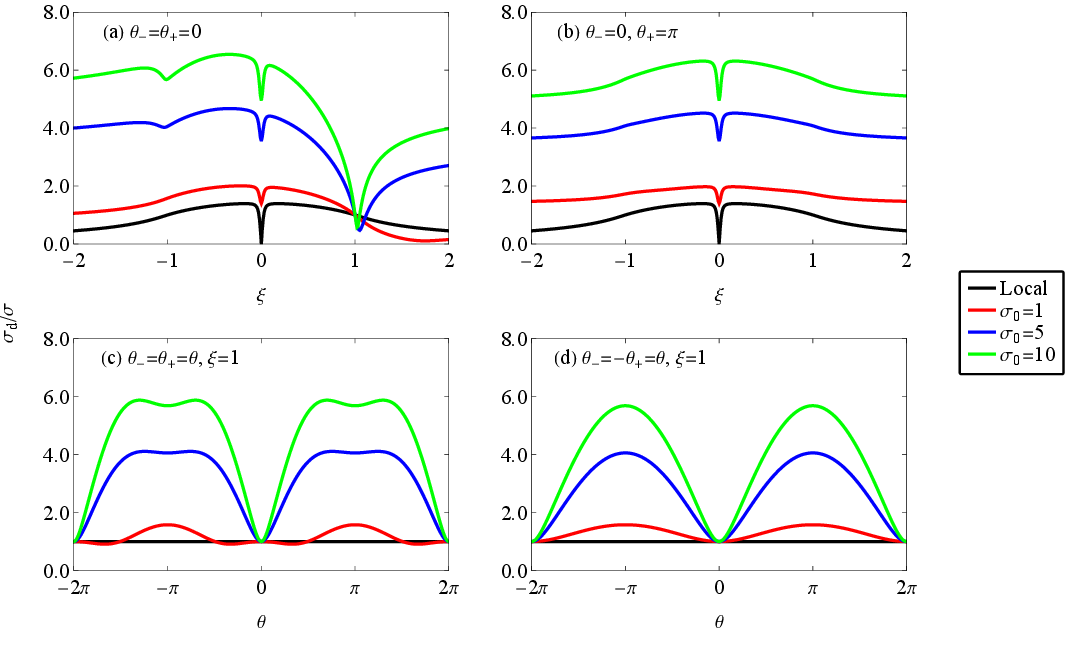}
\caption{Ratio $\sigma_d/\sigma$ of the standard deviations in the defective and homogeneous lattices. (a) and (b) show this ratio as a function of $\xi$ for $\theta_-=\theta_+=0$ and $\theta_-=0$ and $\theta_+=\pi$, respectively. (c) and (d) show the same ratio as a function of $\theta$ for $\theta_-=\theta_+=\theta$ and $\theta_-=-\theta_+=\theta$, respectively, with $\xi=1$. The zero-phase configuration in (a) produces asymmetric spreading for Gaussian initial states, whereas the choice $\theta_-=0$ and $\theta_+=\pi$ in (b) suppresses this asymmetry. Gaussian initial states are more strongly affected by the defect than the fully localized state. The black curve corresponds to the localized initial state $\ket{\psi_0}=\ket{0}$, while the Gaussian initial states have $\sigma_0=1$ (red curve), $5$ (blue curve), and $10$ (green curve). In all panels, the evolution is shown up to $\gamma t=1000$.}
\label{Fig5}
\end{figure*}

Figures \ref{Fig4}(a) and \ref{Fig4}(b) also highlight a pronounced asymmetry for Gaussian initial states for $\theta_-=\theta_+=0$ and a restoration of symmetry when $\theta_-=0$ and $\theta_+=\pi$ are applied. This behavior originates from phase-induced chirality: Localized initial states, whose evolution begins solely at $j=0$, preserve spatial symmetry, whereas Gaussian states distribute amplitude over multiple sites and acquire different phases when propagating from $j=\pm1$ toward the defect. These phase-shifted contributions generate a directional bias in the probability flow. When the phases are $\theta_+=\pi$ and $\theta_-=0$, destructive interference removes this bias, restoring the symmetric profile and demonstrating direct control of state localization via local phase tuning. See the Appendix, where this mechanism is demonstrated analytically on defective $L_5$. Figures \ref{Fig4}(c) and \ref{Fig4}(d) show $P_0$ as a function of the phase $\theta$ for $\theta_-=\theta_+=\theta$ and $\theta_-=-\theta_+=\theta$, respectively, for the time frames $\gamma t=100$ (dashed lines) and $\gamma t=1000$ (solid lines). To show both cases on the same scale, we rescaled $P_0$ by their respective $\gamma t$, finding that $P_0$ vanishes at long times. As a representative example, the insets in Figs.~\ref{Fig4}(c) and \ref{Fig4}(d) show $P_0$ as a function of $\gamma t$ for $\sigma_0=10$ and $\theta=\pi/4$. A power-law fit confirms the asymptotic scaling $P_0\propto(\gamma t)^{-1}$ once the long-time regime is reached. This power-law decay suggests that the defect does not support persistent trapping in this regime, consistent with the known return-probability scaling of one-dimensional CTQWs~\cite{konno2010localization}. Thus, although phase defects can transiently influence state localization, they do not generate long-lived bound states under the parameters explored here. 

Figure~\ref{Fig5} illustrates how the defect parameters $(\xi,\theta_\pm)$ influence transport in CTQWs initialized from Gaussian states with different widths. Figures 5(a) and 5(b) further confirm the chiral effect, showing asymmetric and symmetric profiles of $\sigma_d/\sigma$, respectively, when the phases are $\theta_-=\theta_+=0$ and $\theta_-=0$ and $\theta_+=\pi$. Together with the probability distributions, these results demonstrate direction-dependent propagation for Gaussian initial states under equal phases, consistent with earlier observations~\cite{khalique2021controlled}. They also reveal a striking feature: While a localized state shows a modest enhancement of approximately $39\%$ in $\sigma_d/\sigma$, this improvement grows markedly with $\sigma_0$, reaching values multiple times larger for highly delocalized states. Thus, although delocalized states propagate more slowly on a homogeneous lattice, the presence of a single local defect can substantially boost their transport efficiency—exceeding even the gains observed for a localized walker. Moreover, an equally pronounced enhancement can be achieved via phase engineering. As shown in Figs.~\ref{Fig5}(c) and \ref{Fig5}(d), when the hopping amplitude is fixed at $\xi=1$, tuning only the phases $\theta_\pm$ is sufficient to produce a comparable increase in the spreading rate.

Phase control in CTQWs recently emerged as a powerful mechanism for manipulating quantum interference and steering transport. Engineered phases, combined with suitable graph structures and initial states, can selectively suppress the occupation of target sites without hindering global spreading~\cite{chaves2023why,sett2019zero} and can likewise enhance propagation by coherently modulating interference pathways~\cite{li2015single,ximenes2024parrondo}. The results in Figs.~\ref{Fig4} and~\ref{Fig5} demonstrate these mechanisms in a minimal setting for delocalized states: Phase modulation enables or suppresses directional transport without altering the asymptotic delocalized regime and can even enhance propagation by reshaping interference-driven spreading dynamics. This tunability underscores the role of phases as an accessible and versatile resource for coherent control in quantum walks.

Our results so far indicate that appropriate choices of $(\xi,\theta_\pm)$ can substantially enhance the transport of quantum information by Gaussian states, either symmetrically or with directional bias. Although this highlights a distinctive feature of delocalized wave packets, it also reveals parameter regimes that are unfavorable for propagation. As shown in Fig.~\ref{Fig5}(c), for $\sigma_0=1$ the ratio $\sigma_d/\sigma$ exhibits minima at $\theta=\pm\pi/3,\pm5\pi/3$ for $\xi=1$, corresponding to conditions where $\sigma_d/\sigma<1$. These minima motivate a systematic search for poorer-spreading configurations, which then serve as the individual ``losing'' strategies underpinning the Parrondo-type alternation protocol for a Gaussian state examined in the next section.

\subsection{Parrondo protocol with a Gaussian state}

Parrondo’s effect, originally introduced in game theory, describes situations in which individually losing strategies, when alternated, yield an overall winning outcome~\cite{harmer2002review,shu2014beyond}. In the present setting, a losing strategy corresponds to a CTQW on a defective lattice whose parameters $(\xi,\theta_\pm)$ produce a spreading rate lower than that of the homogeneous case ($\sigma_d/\sigma<1$), whereas a winning strategy corresponds to a configuration with a spreading rate greater than that of the homogeneous case ($\sigma_d/\sigma>1$). Ximenes \textit{et al.} investigated this phenomenon for localized initial states by alternating between two real-valued defect parameters~\cite{ximenes2024parrondo}. Because delocalized states on defective lattices typically outperform the homogeneous lattice, it is first necessary to identify regions of parameter space where their spreading is reduced, so that losing strategies can be appropriately defined.

\begin{figure}[!ht]
\includegraphics[width=\linewidth]{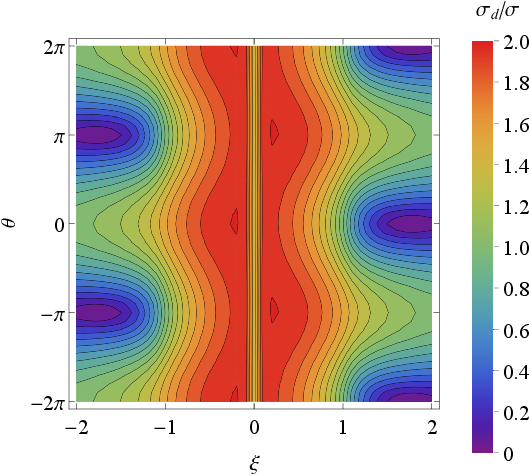}
\caption{Contour plot of the ratio $\sigma_d/\sigma$ as a function of the defect parameters $\xi$ and $\theta$. The simulation was performed for a Gaussian initial state ($\sigma_0=1$) and evolved up to $\gamma t = 1000$.}
\label{Fig6}
\end{figure}

Figure~\ref{Fig6} shows a contour map of the ratio $\sigma_d/\sigma$ as a function of the defect parameters $(\xi,\theta)$ for the configuration $\theta_-=-\theta_+=\theta$, whose behavior closely resembles the case $\theta_-= \theta_+ = \theta$. The color scale makes clear that the spreading efficiency depends sensitively on both the hopping-amplitude ratio $\xi$ and the phase $\theta$. Near the homogeneous limit ($\xi\approx 1$, $\theta\approx 0$), the contour map displays a narrow band of values close to unity (light green region), consistent with $\sigma_d/\sigma\simeq 1$. In contrast, two extended regions of suppressed spreading appear as dark blue lobes centered around the parameters $(\xi\simeq 2,\,\theta\simeq 0,2\pi)$ and $(\xi\simeq -2,\,\theta\simeq \pm\pi)$, where $\sigma_d/\sigma<1$. These minima arise from strong-bond defects that generate reflection-dominated dynamics: Constructive interference between left- and right-moving components traps a significant fraction of the probability density near the defect, thereby inhibiting global spreading. By comparison, a broad yellow-orange region emerges for $-1<\xi<1$ over the full range of $\theta$, indicating $\sigma_d/\sigma>1$ and therefore enhanced spreading. This enhancement arises from destructive interference between left- and right-moving components induced by the reduced defect coupling, which biases transport away from the defect, although a considerable fraction of probability remains weakly trapped. The smooth color transitions surrounding the enhancement and suppression zones show how the boundaries between these regimes shift continuously as $\xi$ departs from unity, demonstrating that the local hopping-amplitude imbalance, together with the phase modulation, jointly shapes the transport behavior. The narrow central vertical region corresponds to $\xi\simeq0$, where the defect bonds are effectively switched off. In this limit, the defect site is decoupled from its nearest neighbors, so the fraction of the initial wave packet located at the defect remains trapped throughout the evolution, producing the vertical stripe observed in Fig.~\ref{Fig6}.

\begin{figure}[!ht]
\includegraphics[width=\linewidth]{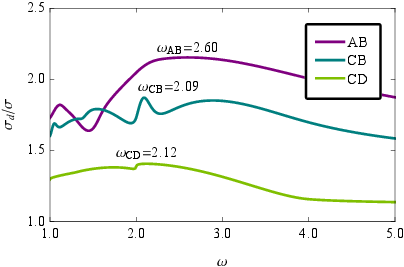}
\caption{Ratio $\sigma_d/\sigma$ as a function of the parameter $\omega$. The value of $\omega$ selected for each alternation protocol is indicated. Simulations were performed for a Gaussian initial state ($\sigma_0 = 1$) and evolved up to $\gamma t = 2000$.}
\label{Fig7}
\end{figure}

After extensive testing, we selected the losing strategies A, B, C, and D listed in Table~\ref{tab.1}. The parameter space was systematically explored to identify representative cases exhibiting distinct dynamical behaviors (strong localization, weak localization, and asymmetric spreading). The selected strategies A–D were chosen as minimal representatives that capture these regimes and define the three Parrondo alternation protocols AB, CB, and CD investigated here. As discussed earlier, each protocol evolves the state in time intervals of $\pi/\omega$, switching between two distinct losing strategies at successive steps. For each protocol, we analyze the dependence on $\omega$ to identify the optimal value that maximizes the spreading enhancement, as shown in Fig.~\ref{Fig7}.

\begin{table}[!ht]
\caption{Defect parameters for each losing strategy (LS) and the corresponding spreading ratio $\sigma_d/\sigma$ at $\gamma t=1000$.}
\label{tab.1}
\begin{ruledtabular}
\begin{tabular}{ccccc}
LS        & $\xi$   & $\theta_-/\pi$ & $\theta_+/\pi$ & $\sigma_d/\sigma$ \\
\hline
A         & $-1.8$  & $-6/5$         & $6/5$          & $0.36$            \\
B         & $1.4$   & $-3/2$         & $3/2$          & $0.91$            \\
C         & $-1.0$  & $11/10$        & $11/10$        & $0.98$            \\
D         & $-1.0$  & $9/10$         & $9/10$         & $0.98$            \\
\hline
\end{tabular}
\end{ruledtabular}
\end{table}

\begin{figure*}[!ht]
\includegraphics[width=\linewidth]{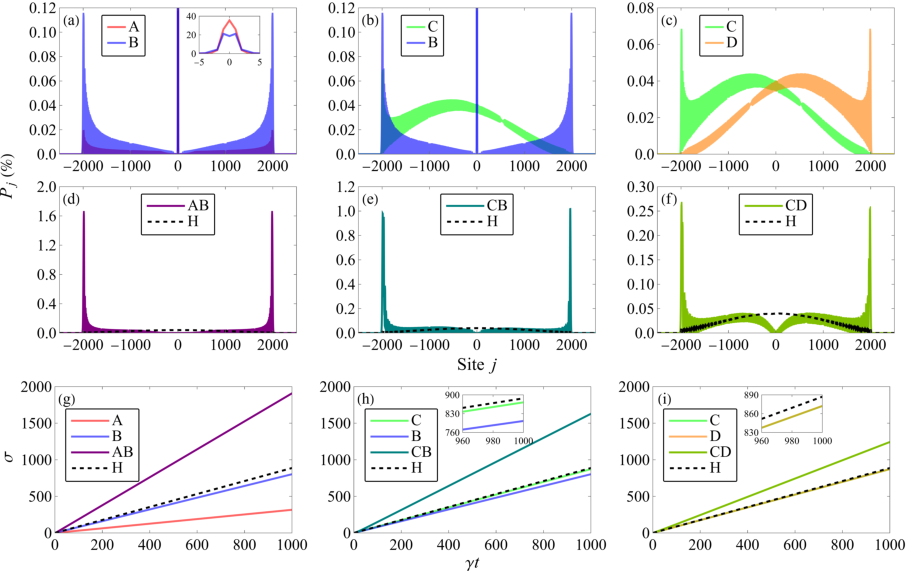}
\caption{Probability distribution $P_j$(a)--(c) for losing strategies A (red curve), B (blue curve), C (green curve), and D (orange curve) and (d)--(f) for the alternation protocols AB (purple curve), CB (dark teal curve), and CD (lime green curve). (g)--(i) show the corresponding standard deviations $\sigma$ as a function of time, comparing each pair of losing strategies with its respective Parrondo alternation protocol. The strategies are defined by the defect parameters: A, $(\xi=-1.8,\,\theta_-=-1.2\pi,\,\theta_+=1.2\pi)$; B, $(\xi=1.4,\,\theta_-=-1.5\pi,\,\theta_+=1.5\pi)$; C, $(\xi=-1.0,\,\theta_\pm=1.1\pi)$; and D, $(\xi=-1.0,\,\theta_\pm=0.9\pi)$. The homogeneous lattice H (black dashed line) is included for reference. The insets in (h) and (i) show that C, B, and D remain below H throughout the evolution; in (i), the curves for C and D are nearly indistinguishable. All simulations were performed for a Gaussian initial state with $\sigma_0=1$ and evolved up to $\gamma t = 1000$.}
\label{Fig8}
\end{figure*}

Figure~\ref{Fig8} displays the transport features of the four losing strategies, which capture distinct dynamical behaviors of CTQWs on defective lattices, together with the outcomes of the three Parrondo alternation protocols constructed from them. Figures~\ref{Fig8}(a)--\ref{Fig8}(c) display the probability distributions $P_j$ for the individual losing strategies. Figure~\ref{Fig8}(a) shows that both strategies A and B produce nearly symmetric profiles with pronounced localization around the defect site. Strategy~A exhibits strong trapping, concentrating approximately $97\%$ of the total probability for $j\in[-4,4]$, whereas B retains about $76\%$ in the same interval, as shown in the inset. Although their global spreadings differ, both cases display severely hindered transport. Case B highlights the importance of jointly examining spreading and localization: Its standard deviation is only about $9\%$ smaller than that of the homogeneous lattice, yet the walker’s transport remains substantially inhibited. More generally, a defect configuration may satisfy $\sigma_d/\sigma > 1$ while still exhibiting noticeable local trapping, showing that enhanced spreading does not imply the absence of localization. The remaining losing strategies, C and D, shown in Figs.~8(b) and 8(c), generate asymmetric (left- and right-biased) distributions with only weak residual localization near the defect. Even in the absence of strong trapping, these strategies still fail to achieve efficient spreading, demonstrating that poor transport does not require pronounced localization.

Figures~\ref{Fig8}(d)--\ref{Fig8}(f) present the probability distributions obtained from alternating the corresponding losing strategies in time. In all three cases, the alternation protocol eliminates the localization present in the individual strategies, yielding broad distributions that exceed the homogeneous benchmark H (black dashed lines). The suppression of trapping arises because each half-evolution step disrupts the constructive interference responsible for localization in the individual strategies, allowing probability to escape from the defect region: The state is effectively ``pushed away'' from the defect, eliminating trapping and promoting outward transport. Beyond merely removing localization, the alternation generates a qualitatively new dynamical regime, which converts two individually poor-spreading evolutions into a markedly enhanced one.

The consequences of this enhancement are quantified in Figs.~\ref{Fig8}(g)--\ref{Fig8}(i), which show the time dependence of the standard deviation $\sigma$ for each pair of losing strategies and for their alternated evolution. In every case, the alternation protocol produces a significantly faster growth of $\sigma$ than either of its constituent losing strategies and even surpasses the homogeneous evolution for long times. The insets in Figs.~\ref{Fig8}(h) and \ref{Fig8}(i) confirm that individual strategies B, C, and D remain below the homogeneous curve throughout the evolution, emphasizing that each strategy is indeed ``losing.'' In contrast, the alternated sequence displays sustained growth with an asymptotic slope larger than that of all individual evolutions. The nearly overlapping curves for C and D in Fig.~\ref{Fig8}(i) further illustrate the robustness of this behavior for strategies with similar dynamical signatures.

Taken together, the results in Fig.~\ref{Fig8} demonstrate that alternating two losing strategies can suppress defect-induced localization, restore long-range transport either symmetrically [Figs.~\ref{Fig8}(d) and \ref{Fig8}(f)] or with directional bias [Fig.~\ref{Fig8}(e)], and produce a spreading rate that exceeds both individual dynamics and the homogeneous quantum walk. This constitutes a clear realization of a Parrondo-type effect for CTQWs initialized from Gaussian states. 

For completeness, we also compared the Parrondo-type protocols with alternations involving representative winning strategies, defined by $\sigma_d/\sigma>1$. We found that, in selected regimes, protocols constructed from losing strategies can perform comparably to, or even outperform, such alternations of winning strategies while simultaneously reducing localization near the defect. Thus, the aim of the present work is not to optimize over the full parameter space $(\xi,\theta_\pm,\omega)$, but rather to demonstrate that alternating two individually poor-spreading dynamics can generate enhanced transport.

\section{Concluding Remarks}\label{sec:4} 

We studied the transport properties of CTQWs initialized in delocalized states on a one-dimensional lattice with a single defect. The defect is implemented through a modified hopping amplitude $\xi$ and complex phases $\theta_-$ and $\theta_+$ applied to the left and right bonds connecting the walker’s central site to its nearest neighbors. We focused on the behavior of Gaussian initial states with different widths in comparison with the fully localized case. Like localized states, Gaussian wave packets are sensitive to the defect strength, but they additionally display a phase-induced chirality absent in the localized case. We further showed that their dynamics reflect a competition between weak- and strong-coupling localization mechanisms, each modulated by the applied phases.

Our results show that suitable choices of $(\xi,\theta_\pm)$ can significantly enhance transport for Gaussian initial states, even though delocalization typically slows spreading on homogeneous lattices. In particular, phase engineering enables fine control over directional propagation, allows suppression or restoration of spatial symmetry, and can boost the asymptotic spreading rate without inducing persistent localization. Conversely, we identified broad regions of parameter space where the standard deviation is reduced or where noticeable trapping persists despite $\sigma_d/\sigma > 1$, demonstrating that spreading efficiency and localization must be examined jointly when characterizing transport.

Mapping the $(\xi,\theta)$ parameter space revealed extended regions of enhanced and suppressed spreading governed by the interplay between hopping-amplitude imbalance and phase modulation. Regions of poor transport, identified through $\sigma_d/\sigma < 1$ for $\sigma_0=1$, provided the foundation for defining losing strategies in a Parrondo-type framework. Using these configurations, we constructed three alternation protocols, each formed by switching between two individually losing strategies at time intervals of $\pi/\omega$. For appropriately chosen $\omega$, the alternation protocols suppress localization, redistribute trapped probability, and achieve spreading rates that surpass those of both constituent strategies and the homogeneous lattice. This behavior underscores how temporal alternation between static defects can generate constructive interference patterns that enhance global transport.

Overall, our findings demonstrate that defect engineering, phase control, and temporal alternation protocols constitute accessible resources for manipulating quantum states in CTQWs and improving their transport properties. The interplay of these mechanisms offers new avenues for steering quantum transport, designing quantum-walk-based experiments, and developing controllable quantum network architectures. Extensions to multidefect systems, higher-dimensional geometries, and explicitly time-dependent control protocols provide natural directions for future work.

\section*{Acknowledgments}
This work was supported by Conselho Nacional de Desenvolvimento Científico e Tecnológico (CNPq) through Grant No. \mbox{409673/2022-6}, by the National Institute for Science and Technology of Quantum Information (INCT-IQ) under Grant No. \mbox{465469/2014-0} and by the National Institute of Science and Technology for Applied Quantum Computing (INCT-CQA) under Process No. \mbox{408884/2024-0}. Computational resources were provided by the Centro Nacional de Processamento de Alto Desempenho em São Paulo (CENAPAD-SP). E.P.M.A. thanks J. Longo for her careful reading of the manuscript.

\section*{Data Availability}
There are no publicly available research data or software supporting this manuscript. Requests for further information or data should be sent to the authors.

\renewcommand{\appendixname}{APPENDIX}
\appendix*
\section{SYMMETRIES IN CTQWS ON DEFECTIVE $L_5$}

\setcounter{equation}{0}
\renewcommand{\theequation}{A\arabic{equation}}

We present an analytical example of CTQWs initialized from both localized and delocalized states on the five-site graph $L_5$, whose sites are labeled by $j \in \{-2,-1,0,1,2\}$. The defect is introduced through complex hopping amplitudes of the form $\xi e^{i\theta_\pm}$ connecting the central site $j=0$ to its nearest neighbors $j=\pm1$. Our analysis aims to highlight the emergence of transport bias and its dependence on the defect strength $\xi$ and phases $\theta_\pm$ for delocalized initial states, as well as its absence for localized initial states. The Hamiltonian reads
\begin{equation}
H=\begin{bmatrix}
0 & -1 & 0 & 0 & 0  \\[2pt]
-1 & 0 & -\xi e^{-i\theta_-} & 0 & 0  \\[2pt]
0 & -\xi e^{i\theta_-} & 0 & -\xi e^{-i\theta_+} & 0 \\[2pt]
0 & 0 & -\xi e^{i\theta_+} & 0 & -1 \\[2pt]
0 & 0 & 0 & -1 & 0
\end{bmatrix}.
\label{HamL5}
\end{equation}
We first consider an initially localized state at the central site, $\ket{\psi_0} = \ket{0}$. The probability of occupation of site $j$ at time $t$ is given by
\begin{equation}
P_j(t) = \left| \braket{j|\psi(t)} \right|^2,
\label{prob}
\end{equation}
and the time-evolved state reads 
\begin{equation}
\ket{\psi(t)} = \sum_{k} e^{-i\lambda_k t} \braket{\phi_k|\psi_0} \ket{\phi_k},
\label{evolvedpsi}
\end{equation}
where $\lambda_k$ and $\ket{\phi_k}$ denote the eigenvalues and corresponding eigenvectors of the Hamiltonian in Eq.~\eqref{HamL5}. Diagonalizing the Hamiltonian yields the eigenvalues
\begin{equation}
\lambda = \{-1,\, 1,\, 0,\, -\beta,\, \beta\},
\end{equation}
where $\beta = \sqrt{2\xi^2 + 1}$. The corresponding eigenvectors yield the following explicit form of the evolved state:
\begin{align}
\ket{\psi(t)}
&=
\beta^{-2}
\begin{bmatrix}
e^{-i\theta_-}\xi\!\left[\cos(\beta t)-1\right] \\[2pt]
ie^{-i\theta_-}\beta\xi\sin(\beta t) \\[2pt]
2\xi^{2}\cos(\beta t)+1 \\[2pt]
ie^{i\theta_+}\beta\xi\sin(\beta t) \\[2pt]
e^{i\theta_+}\xi\!\left[\cos(\beta t)-1\right]
\end{bmatrix},
\end{align}
which yields the probabilities of finding the particle at each site,
\begin{align}
P_{\pm 2}(t) &= \frac{4\xi^2}{\beta^4}\,\sin^4\!\left(\frac{\beta t}{2}\right),\\
P_{\pm 1}(t) &= \frac{\xi^2}{\beta^2}\,\sin^2(\beta t),\\
P_0(t) &= \frac{\left[2\xi^2 \cos(\beta t)+1\right]^2}{\beta^4}.
\end{align}
Thus, the probability distribution is invariant under inversion of the spatial index, $P_j(t)=P_{-j}(t)$, and also under $\xi \to -\xi$. Although the phases explicitly appear in the amplitudes, they cancel in the probabilities, ensuring the absence of directional bias in the dynamics. This symmetry is further reflected in the mean position, $\braket{j}$, which vanishes identically for all times. The standard deviation is given by
\begin{equation}
\sigma(t) = \frac{2|\xi|}{\beta^2}\sqrt{8\sin^4\!\left(\frac{\beta t}{2}\right) + \frac{\beta^2}{2}\sin^2(\beta t)},
\end{equation}
which is also invariant under inversion of $\xi$. 

These results show that, for an initially localized state, the CTQW preserves a symmetric probability distribution regardless of the presence of complex hopping amplitudes or variations in the defect strength. This establishes that a localized initial state does not lead to chiral or directional transport in this scenario.

We now consider a delocalized initial state given by a uniformly distributed state over sites $j=\{-1,0,1\}$, which can be written as
\begin{equation}
\ket{\psi_0}=\frac{1}{\sqrt{3}}(\ket{-1}+\ket{0}+\ket{1}).
\end{equation}
Following the same procedure as above, we obtain
\begin{widetext}
\begin{align}
\ket{\psi(t)} &= \frac{1}{2\sqrt{3} \beta ^2}
\begin{bmatrix}
2\xi e^{-i\theta_-}\left(\cos\left(\beta t\right)-1\right)
- i\beta e^{-i\left(\theta_-+\theta_+\right)}\left(\beta\sin t - \sin\left(\beta t\right)\right)
+ i\beta\left(\beta\sin t + \sin\left(\beta t\right)\right)\\[2pt]
\beta\Big[
\beta\left(\cos t + \cos\left(\beta t\right)\right)
+ \beta e^{-i\left(\theta_-+\theta_+\right)}\left(\cos\left(\beta t\right)-\cos t\right)
+ 2i\xi\sin\left(\beta t\right)e^{-i\theta_-}\Big]\\[2pt]
2\Big[
1 + 2\xi^2\cos\left(\beta t\right)
+ i\beta\xi\sin\left(\beta t\right) \left(e^{i\theta_-} + e^{-i\theta_+}\right)\Big]\\[2pt]
\beta\Big[\beta\left(\cos t + \cos\left(\beta t\right)\right)
- \beta e^{i\left(\theta_-+\theta_+\right)}\left(\cos t - \cos\left(\beta t\right)\right)
+ 2i\xi\sin\left(\beta t\right)e^{i\theta_+}\Big]\\[2pt]
2\xi e^{i\theta_+}\left(\cos\left(\beta t\right)-1\right)
- i\beta e^{i\left(\theta_-+\theta_+\right)}\left(\beta\sin t - \sin\left(\beta t\right)\right)
+ i\beta\left(\beta\sin t + \sin\left(\beta t\right)\right)
\end{bmatrix},
\end{align}
and the mean position of the walker gives
\begin{equation}
\braket{j}=-\frac{4 \xi}{3 \beta ^2} \sin ^2\left(\frac{\beta  t}{2}\right) \left[\beta  \cos (t) \cot \left(\frac{\beta  t}{2}\right)+2 \sin (t)\right]\left(\sin \theta_-+\sin \theta _+\right).
\end{equation}
\end{widetext}
For equal phases $\theta_-=\theta_+=\theta$, the dynamics becomes biased whenever $\theta\neq k\pi$ ($k\in\mathbb{Z}$) since $\braket{j}$ is proportional to $2\sin\theta$. For $\theta=k\pi$, this contribution vanishes, and the walk remains unbiased. In contrast, for $\theta_-=-\theta_+=\theta$, one obtains $\braket{j}=0$ for any $\theta$.

Last, let us consider the special case $\theta_-=0$ and $\theta_+=\pi$; we have
\begin{align}
\ket{\psi(t)}
&=
\frac{1}{\sqrt{3}\,\beta^{2}}
\begin{bmatrix}
\xi\!\left[\cos(\beta t)-1\right]+i\beta^{2}\sin t \\[2pt]
\beta\!\left[\beta\cos t+i\xi\sin(\beta t)\right] \\[2pt]
2\xi^{2}\cos(\beta t)+1 \\[2pt]
\beta\!\left[\beta\cos t-i\xi\sin(\beta t)\right] \\[2pt]
-\xi\!\left[\cos(\beta t)-1\right]+i\beta^{2}\sin t
\end{bmatrix},
\end{align}
which gives the following probabilities:
\begin{align}
P_{\pm 2}(t) &= \frac{1}{3} \left[\frac{\xi ^2 (\cos (\beta  t)-1)^2}{\beta ^4}\!+\!\sin ^2(t)\right],\\
P_{\pm 1}(t) &= \frac{1}{3} \left[\frac{\xi ^2 \sin ^2(\beta  t)}{\beta ^2}\!+\!\cos ^2(t)\right],\\
P_0(t) &= \frac{\left(2 \xi ^2 \cos (\beta  t)\!+\!1\right)^2}{3 \beta ^4},
\end{align}
where the probability distribution is invariant under inversion of both $j$ and $\xi$, as in the localized case. Although Gaussian and truncated uniform states differ, the mechanism underlying the $\xi$ symmetry and the unbiased dynamics for $\theta_-=0$ and $\theta_+=\pi$ is common to both cases. This suggests that nonzero amplitudes at sites adjacent to $j=0$, which allow probability flux from $j=\pm1$ toward the defect from the outset, are sufficient to induce phase-controlled asymmetry in delocalized states.

\end{document}